\documentclass[aps,twocolumn,superscriptaddress,floatfix,amsmath,amssymb]{revtex4}
\usepackage{graphicx}
\usepackage{amssymb}
\newcommand{\on}[2]{\mathop{\null#2}\limits^{#1}} 
\newcommand{\oover}[1]{\on{\circ}{#1}}

\begin{document}

\title{Viscoelasticity and Stokes-Einstein relation in repulsive and attractive colloidal glasses}

\author{Antonio M. Puertas}
\affiliation{Group of Complex Fluids Physics, Departamento de F\'{\i}sica Aplicada, Universidad de Almer\'{\i}a, 04120 Almer\'{\i}a, Andaluc\'{\i}a, SPAIN}
\author{Cristiano De Michele}
\affiliation{Dipartimento di Fisica, Istituto Nazionale per la
Fisica della Materia, and CNR-INFM-SOFT,
Universit\`a di Roma La Sapienza, Piazzale Aldo
Moro~2, I-00185, Roma, Italy}
\author{Francesco Sciortino}
\affiliation{Dipartimento di Fisica, Istituto Nazionale per la
Fisica della Materia, and CNR-INFM-SOFT,
Universit\`a di Roma La Sapienza, Piazzale Aldo
Moro~2, I-00185, Roma, Italy}
\author{Piero Tartaglia}
\affiliation{Dipartimento di Fisica, Istituto Nazionale per la
Fisica della Materia, and CNR-INFM-SMC,
Universit\`a di Roma La Sapienza, Piazzale Aldo
Moro~2, I-00185, Roma, Italy}
\author{Emanuela Zaccarelli}
\affiliation{Dipartimento di Fisica, Istituto Nazionale per la
Fisica della Materia, and CNR-INFM-SOFT,
Universit\`a di Roma La Sapienza, Piazzale Aldo
Moro~2, I-00185, Roma, Italy}

\date{\today}

\begin{abstract}
We report a numerical investigation of the visco-elastic behavior in
models for steric repulsive and short-range attractive colloidal
suspensions, along different paths in the attraction-strength vs
packing fraction plane.  More specifically, we study the behavior of
the viscosity (and its frequency dependence) on approaching the
repulsive glass, the attractive glass and in the re-entrant region
where viscosity shows a non monotonic behavior on increasing
attraction strength.  On approaching the glass lines, the increase of
the viscosity is consistent with a power-law divergence with the same
exponent and critical packing fraction previously obtained for the
divergence of the density fluctuations.  Based on mode-coupling
calculations, we associate the increase of the viscosity with specific
contributions from different length scales. We also show that the
results are independent on the microscopic dynamics by comparing
newtonian and brownian simulations for the same model. Finally we
evaluate the Stokes-Einstein relation approaching both glass
transitions, finding a clear breakdown which is particularly strong
for the case of the attractive glass.
\end{abstract}


\maketitle

\section{Introduction}
Understanding dynamic arrest in colloidal system is crucial in
disparate technological applications (e.g. food industry\cite{Mez05a},
biomaterials\cite{Sear06}, painting).  Development of basic science
also requires a deeper understanding of the different routes and
mechanisms leading to dynamic arrest (glasses and
gels)\cite{Tra04a,Cip05a,Sci05a,Zacca07}. In this respect, model colloidal
systems are playing a very important role due to their versatility.
It is indeed possible to tailor the shape, size and structure of the
colloidal particles making it possible to design specific colloidal
interaction potentials\cite{Yeth03}. Furthermore, accurate
experimental methods are now available for investigating the structure
and the dynamics of colloids even at the single particle
level\cite{Weeks07}.  Unexpected novel behaviors regarding the glass
transition have been theoretically
predicted\cite{Fab99a,Ber99a,Daw00a,Gotzesperl,sperl} and
experimentally
observed\cite{Pha02a,Eck02a,Chen03a,Pha04a,Gra04a,Nara2006PRL} in the
cases in which colloidal particles interact, beside the hard-core, via
a short-range attractive interaction potentials (when the attraction
range is about one tenth of the particle diameter or less).  The
predictions, based on application of the mode coupling theory for
supercooled liquids (MCT)\cite{goetze} suggest that the standard
packing-driven hard-sphere glass transition transforms --
discontinuously in some cases -- into a novel-type of glass transition
driven by the short-range attraction.  The competition between the two
different arrest mechanisms introduces slow-dynamics features which
are not commonly observed in molecular and atomic systems.
Experiments on solutions of (hard-sphere like) colloidal particles
(either PMMA or polystyrene micronetwork spheres) in the presence of
small non-adsorbing polymers \cite{Pha02a,Eck02a,Pha04a} have shown
that there exists a window of polymer densities in which the mobility
of the colloidal particles has a maximum for a finite value of polymer
concentration.  Moreover, for small and large polymer concentrations,
the strength of the $\alpha$-relaxation (the non-ergodicity parameter)
is found to be very different, suggesting that indeed the
visco-elastic response of the repulsive and attractive glass will also
be significantly different.  Molecular dynamics simulations of
short-ranged models\cite{Pue02a,Fof02a,Zac02a,Pue03a} have confirmed
the picture resulting from the theoretical predictions and validated
by the experiments. A recent review can help summarizing the
experimental and numerical studies in short-range attractive
colloids\cite{Sci05a}.

The numerical results have been so far mostly limited to the study of
self and collective properties of the density fluctuations.  Despite
the strong link with experiments and the relevance to industrial
applications, the numerical evaluation of the viscosity, $\eta$, and
viscoelastic properties $\tilde \eta(\omega)$ have lagged behind,
since significant computational effort is requested for accurate
calculation of $\tilde \eta(\omega)$, even more for states close to
dynamical arrest.  Experimentally, measurements of $\eta$ close to the
repulsive hard-sphere glass transition show an apparent divergence,
but there is no consensus on the functional form describing such
increase\cite{cheng02,Fuc03a}. For colloidal gels, a power law
divergence has been reported in connection to the gel transition
\cite{shah03}.  Theoretically, MCT predicts an asymptotic power law
divergence, with identical exponent, of all dynamical quantities with
the distance from the transition, and hence $\eta$, the time scale of
the density fluctuations $\tau$ and the inverse of the self diffusion
coefficient $1/D_0$ should diverge with the same critical parameters.

In this article, we attempt a characterization of the viscoelastic
properties of two different short-range attractive potentials (a
polydisperse Asakura-Osawa and a square-well) along three different
paths in the attraction strength-packing fraction plane, which
allow us to access both the repulsion driven and attraction driven
glass transitions with both systems.  We show the divergence of the
viscosity, as well as the diffusion coefficient or structural
relaxation time, as the repulsive and attractive glasses are
approached. At high density, the isochoric path shows the reentrant
glass; the viscosity increases about three orders of magnitude upon either
increasing or decreasing the strength of attraction.

The article is organized as follow: in Sec.~II we introduce the
numerical models and describe the methods to calculate the
viscosity. In Sec.~III we describe the paths investigated and provide
some background information on the behavior of the diffusion and
collective density fluctuations along these paths. In Sec.~IV we
discuss the observed behavior of the viscosity on approaching the
repulsive and the attractive glass lines. In Sec.~V, guided by
theoretical MCT predictions for the viscosity, we provide evidence
that the visco-elastic behavior close to the two different glass lines
is controlled by density fluctuation of different wavelength. Finally
in Sec. VI we report a study of the density and attraction strength
dependence of the Stoke-Einstein relation.

\section{Numerical Simulations}

\subsection{Model A: Square well and Hard Sphere Binary Mixture}
We perform Molecular Dynamics (MD) simulations of a 50:50 binary
mixture of 700 particles of mass $m$ with diameters $\sigma_{AA}=1.2$
and $\sigma_{BB}=1$ (setting the unit of length).  The particles
interact through a hard core repulsion complemented by a narrow square
well (SW) pair potential.   The hard core repulsion
for the $AB$ interaction occurs at a distance
$\sigma_{AB}=(\sigma_{AA} + \sigma_{BB})/2$.  The SW potential is,
\begin{equation}
V_{SW}(r)= \begin{cases}
        ~~\infty ~~~~~~~\hspace{0.8 mm} r<\sigma_{ij}   \\ 
        -u_0  ~~~~~~~  \sigma_{ij}<r<\sigma_{ij}+\Delta_{ij}\\ 
        ~~~0      ~~~~~~~~ r>\sigma_{ij}+\Delta_{ij}
\end{cases}
\end{equation}
where $r$ is the distance between particles of types $i,j=A,B$, the
depth of the well $u_0$ is set to $1$ and the widths $\Delta_{ij}$ are
such that $\Delta_{ij}/(\sigma_{ij}+\Delta_{ij})=0.03$.  Temperature
$T$ is measured in units of $u_0$ ($k_B=1$), the attraction strength
$\Gamma=1/T$, time $t$ in $\sigma_{BB}(m/u_0)^{1/2}$.  The use of a
binary mixture allows us to suppress crystallization at high packing
fraction $\phi=(\rho_A\sigma_A^3+\rho_B\sigma_B^3)\cdot \pi/6$, where
$\rho_i=N_i/L^3$, $L$ being the box size and $N_i$ the number of
particles for each species.  The system undergoes phase separation
into a gas and a liquid for large attraction strength in a wide range
of packing fractions~\cite{Zaccapri}: the critical point is located
roughly at $\Gamma_c\approx 3.33$ and $\phi_c\approx 0.27$ (the latter
is estimated from the Noro-Frenkel scaling\cite{Nor00aJCP} invariance
close to the Baxter limit\cite{Mil04aJCP}).  Previous
studies\cite{Zac02a,Sci03a,Zaccapri} of the same model allowed us to
locate the dynamical arrest line and the spinodal curve.  The
`numerical' glass line was determined by extrapolation via a power-law
fitting of the normalized diffusion coefficient $D/D_0$,
i.e. $D/D_0\sim (\phi-\phi_g)^\gamma$~\cite{Sci03a}, where
$D_0=\Gamma^{1/2}$ .  This study was complemented by the calculation
of the MCT glass lines for the same model. Hence, a bilinear
transformation of $\phi$ and $T$ was used to to superimpose the
theoretical onto the numerical glass line.

We also study, as discussed below, the same 50:50 binary mixture of
700 particles, with the same $\sigma_{AA},\sigma_{BB},\sigma_{AB}$
above, but interacting simply as hard spheres, for which the potential
reads,
\begin{equation}
V_{HS}(r)=
\begin{cases}
        ~~\infty ~~~~~~~~r<\sigma _{ij}   \\ 
        ~~~0      ~~~~~~~~\hspace{0.8 mm} r>\sigma_{ij}.
\end{cases}
\end{equation}

For Newtonian dynamics (ND) simulations, we used a standard event-driven (ED)
algorithm\cite{rapaport}.
We also perform Brownian Dynamics (BD) simulations of the same model,
to ensure the independence of the viscoelastic calculations on the
microscopic dynamics. 
For BD simulations we exploit a recently developed  \cite{FoffiDemPRL} 
BD algorithm, which we 
shortly describe below. For a more extensive discussion 
we invite the reader to consult Ref. \cite{DemBDalgo}.
 
If the position Langevin equation is considered, i.e.:
\begin{equation}
\dot {\mathbf r_i} (t) = \frac{D_0}{k_B T} {\mathbf f}_i(t) +
{\oover{\mathbf r}}_i(t),
\label{Eq:langevin}
\end{equation}
where $ {\mathbf r_i} (t) $ is the position of particle $i$, $D_0$ is
the short-time (bare) diffusion coefficient, ${\mathbf f}_i(t)$ is the
total force acting on the particle, $ {\oover{\mathbf r}}_i(t) $ a
random thermal noise satisfying $< {\oover{\mathbf r}}_i(t)\cdot
{\oover{\mathbf r}}_i(0)> = 6D_0 \delta(t)$.
The BD integration scheme of
Eq. \ref{Eq:langevin} can be schematized as follow:
\begin{itemize}
\item[(i)] every $t_n=n\Delta t$ ($n$ integer) extract velocities $\vec v_i$ according to a Maxwellian distribution of variance $\sqrt{k_BT/m}$; 
\item[(ii)] evolve the system between $t_n$ and $t_n+\Delta t$ according to the laws of ballistic motion (performing standard ED molecular dynamics).
\end{itemize}
In other words, Gaussian particle displacements $\Delta \vec{r_i} =
\vec{v_i} \Delta t$ are extracted according to $\langle\Delta
\vec{r_i}^2\rangle = 6 D_0 \Delta t$ and between two velocities
extractions, standard ED dynamics is applied.

The present binary mixture model allows us to study the viscoelastic
properties within the reentrant liquid region, enclosed by the nearby
attractive and repulsive glass transitions.  On the other hand, due to
phase separation, it does not allow us to approach the attractive
glass line at moderate density.  Hence we will study $V_{HS}$ for
varying $\phi$ (Path 1A in Fig.~\ref{fig:paths}) and $V_{SW}$ at
fixed $\phi=0.58$ on varying $T$ (Path 3 in Fig.~\ref{fig:paths}).

\subsection{Model B: Asakura-Oosawa Polydisperse System}

We also study an interaction potential based on the
Asakura-Oosawa model to make a direct link with experiments in
colloid-polymer mixtures. A polydisperse system, comprised of 1000
particles, is simulated with the standard velocity Verlet algorithm
for Newtonian Dynamics in the canonical ensemble, which requires a
continuous differentiable potential. To this end, a soft core was used
instead of the hard core in Model A:

\begin{equation}
V_{sc}(r)\:=\:\left(\sigma_{ij}\right/r)^{36}
\end{equation}
\noindent where $\sigma_{ij}=(\sigma_i+\sigma_j)/2$, with $\sigma_i$
the diameter of particle $i$. Diameters where distributed according to
the flat distribution $[\sigma-\delta,\sigma+\delta]$ with $\sigma$
the mean diameter and $\delta=0.1\sigma$. The short-range attraction
between particles is given by the Asakura-Oosawa model for
polydisperse systems:
\[ V_{AO}(r) \:=\: -k_BT \phi_p \left\{\left[\left(\bar{\eta}+1\right)^3 -\frac{3r}{2\xi} \left(\bar{\eta}+1\right)^2+\frac{r^3}{2\xi^3}\right]+ \right.\]
\begin{equation}\label{pot} 
\left.+\frac{3\xi}{8r} \left(\eta_1-\eta_2\right)^2 \left[\left(\bar{\eta}+1\right) -\frac{r}{\xi} \right]^2\right\}
\end{equation}

\noindent for $\sigma_{12} \leq r \leq \sigma_{12}+\xi)$ and $0$ for
larger distances; $\eta_i=\sigma_i/\xi$,
$\bar{\eta}=(\eta_1+\eta_2)/2$, and $\phi_p$ is the volume fraction of
the polymer. The range of the interaction, $\xi$, is the polymer size,
and its strength is proportional to $\phi_p$, the concentration of
ideal polymers. To ensure that the interaction potential
$V_{sc}+V_{AO}$ has its minimum at $\sigma_{12}$, the Asakura-Oosawa
potential is connected analytically to a parabola at
$\sigma_{12}+\xi/10$ \cite{puertas03}. For average particles,
$\sigma_1=\sigma_2=\sigma$, the attraction strength of the
Asakura-Oosawa potential is given by $V_{min}=-k_BT \phi_p (3/2 \eta
+1)$, which for $\xi=0.1$, is $V_{min}=-16k_BT\phi_p$.

Because the attractive glass transition occurs inside the liquid-gas
spinodal, it cannot be accessed directly from the fluid with this
potential. Thus, we have added a long range repulsive barrier to the
interaction potential that destabilizes a macroscopic separation into
two fluid phases. The barrier is given by:

\begin{equation}
V_{bar}(r)\:=\:k_BT\left\{\left(\frac{r-r_1}{r_0-r_1}\right)^4-2\left( \frac{r-r_1}{r_0-r_1}\right)^2+1\right\}
\end{equation}

\noindent for $r_0\leq r \leq r_2$ and zero otherwise, with
$r_1=(r_2+r_0)/2$. The limits of the barrier were set to
$r_0=\sigma_{12}+\xi$, and $r_2=2\sigma$, and its height is $1
k_BT$. The barrier raises the energy of a dense phase, so that
liquid-gas separation is suppressed. The resulting total interaction,
\begin{equation}
V_{tot}(r)=V_{sc}(t)+V_{AO}(r)+V_{bar}(r)
\end{equation}
 is analytical everywhere and allows straightforward integration of
the equations of motion.

This model allows us to study the viscoelastic properties of the fluid
close to the attraction driven glass transition at moderate density,
i.e. far from the high order singularity. We will use this system to
approach the repulsive glass with increasing $\phi_c$ at $\phi_p=0$,
hence using simply $V_{sc}$ (Path 1B in Fig.~\ref{fig:paths}), as well
as to study the attractive glass at moderate density $\phi_c=0.40$
(Path 2 in Fig.~\ref{fig:paths}) by using $V_{tot}$.

\subsection{Computation of viscosity}

The shear viscosity $\eta$ is given by the Green-Kubo relation:

\begin{equation}
\eta\: \equiv \:\int_0^{\infty} dt\,C_{\sigma\sigma}(t)
\:=\:\frac{\beta}{3V}\int_0^{\infty} dt\,\sum_{\alpha < \beta} \langle
\sigma^{\alpha\beta}(t) \sigma^{\alpha\beta}(0) \rangle,
\label{eq:eta}
\end{equation}

\noindent which expresses $\eta$ as the integral of the correlation
function of the non-diagonal terms of the microscopic stress tensor,
$\sigma^{\alpha\beta}\:=\:\sum_{i=1}^Nm
v_{i\alpha}v_{i\beta}\,-\,\sum_{i<j}^N \frac{r_{ij\alpha}
r_{ij\beta}}{r_{ij}} V'(r_{ij})$, where $V$ is the volume of the
simulation box, $v_{i\alpha}$ is the $\alpha$-th component of the
velocity of particle $i$, and $V'$ is the derivative of the total
potential. $\langle ... \rangle$ indicates an average over initial
conditions. However, from the computational point of view it is more
convenient to use the Einstein relation,
\begin{equation}
\eta\:=\:\lim_{t \rightarrow \infty} \eta(t)\:=\:\frac{\beta}{6V}
\lim_{t\rightarrow \infty} \frac{1}{t} \langle \Delta A(t)^2 \rangle,
\label{eq:einstein}
\end{equation}
\noindent where $\Delta A(t)$ is the integral from $0$ to $t$ of the
three off-diagonal terms of the stress tensor,
\begin{equation}
\Delta A(t)=A(s+t)-A(s)=\int_s^{s+t} \sum_{\alpha < \beta}
\sigma^{\alpha\beta}(s') ds'
\label{eq:At}
\end{equation}
Using Eq.\ref{eq:einstein} is analogous to the calculation of the
diffusion coefficient as the long time slope of the mean squared
displacement.

For discontinuous potentials (hard cores or square wells), equation
\ref{eq:einstein} can still be used\cite{alder} despite the impulsive character
of the interactions. In this case, 
\begin{eqnarray}
[\Delta A(t)]_{HS,SW}=\sum_{collisions}\sum_{\alpha\neq \beta} 
[(m\sum_{i=1}^N v_{i \alpha} v_{i\beta})\tau_t+\nonumber\\
m(x_{k \alpha}-x_{l \alpha})(v_{k \beta}^{after}-v_{l \beta}^{before})]
\label{eq:HSeta}
\end{eqnarray}
where $\tau_t$ is the time elapsed from the previous collision, $k$
and $l$ are the two colliding particles, $x_{k \alpha}$ is the
position of particle $k$ in direction $\alpha$, and $(v_{k
\beta}^{after}-v_{l \beta}^{before})$ is the momentum change in
direction $\beta$ of particle $k$ due to the collision with particle
$l$.  We have not attempted to numerically recover $C_{\sigma
\sigma}(t)$ from $ \Delta A(t)$.

\subsection{Units}
For both studied models we report states in the packing fraction
vs. attraction strength plane ($\phi_c-\Gamma$). For Model A, the
attraction strength is given by the inverse temperature (for HS
temperature is irrelevant and is set equal to 1), whereas for Model B,
$\Gamma=-V_{min}$. Distances are measured using $\sigma_{BB}$ for
model A and the mean diameter, $\sigma$ for model B, while the
particle mass, $m$, is always set to one. The stress correlation
function is measured in units of $k_BT/\sigma^3$, and time in units of
$(\sigma^2 m/k_BT)^{1/2}$. The viscosity is thus given in
$(mk_BT)^{1/2}/\sigma^2$. For the integration of the equations of
motion in model B, the time step was set to $\delta t=0.0025/\sqrt{3}$.

\section{Description of paths, transition, fits, exponents}

\begin{figure}[h] 
\includegraphics[width=0.5\textwidth]{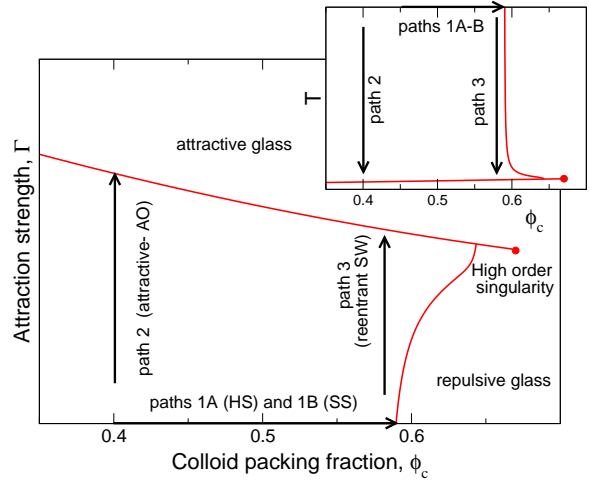}
\caption{Schematic phase diagram showing the attraction and repulsion
driven glasses and the three paths followed in this work. Note that
path 1 (infinite temperature limit) is studied within both
models. The inset shows the three paths in the 
temperature-packing fraction representation.}
\label{fig:paths}
\end{figure}

Using the models presented above, we numerically study the following
paths schematized in Fig.~\ref{fig:paths}:

{\sl Path 1}: The zero-attraction case for both models, i.e. the 
hard- and the soft sphere models. The two models are not
identical along this path because (i) the Asakura-Oosawa model has a
soft repulsion (although the $r^{-36}$-core is quite hard and no
important effects are expected \cite{melrose92}) and more importantly
{\sl ii}) the size distributions are different: bimodal in model A
vs. continuous in model B. Model B has been studied previously along this path monitoring the self-diffusion and the density correlation functions\cite{voightmann04}. The glass transition points and the exponents controlling the power-law divergence of the structural relaxation time
scale, $\gamma_{\tau}$, and the diffusion coefficient, $\gamma_D$, as well as 
the von Schweidler exponent $b$ (which provides a measure of
the slow-decay of the density correlation function), are shown in
Table~\ref{table_RG} for both systems. The difference in the critical packing fractions
can be attributed to the different size distributions of the two models. The
exponents $\gamma_{\tau}$ and $\gamma_D$, on the other hand, are very similar in both models.

\begin{table}
\begin{tabular}{l|cccc}
 & $\phi_c^G$ & $b$ & $\gamma_{\tau}$ & $\gamma_D$ \\ \hline 
Model A: $V_{HS}$ & $0.584$ & $0.51$ & $2.75$ & $2.17$ \\
Model B: $V_{sc}$ & $0.594$ & $0.53$ & $2.72$ & $2.02$ \\
\end{tabular}
\caption{Glass transition point $\phi_c^G$, von
Schweidler exponent $b$, and divergence exponents of the characteristic time of the decay of density fluctuations $\gamma_{\tau}$ and of the diffusion coefficient
$\gamma_D$ for models A and B in the absence of attraction, i.e. respectively
$V_{HS}$ and $V_{sc}$, along path 1.}
\label{table_RG}
\end{table} 

{\sl Path 2}: Approaching the attractive glass. This path is studied
with model B, for which the liquid-gas transition is destabilized and the
glass transition can be approached from the fluid. This path has been
studied previously monitoring the density correlation functions
\cite{puertas03,puertas05} and the viscosity \cite{Pue05a}, and
the glass transition is found for $\Gamma^G=9.099$; the associated von Schweidler and critical exponents are given in Table \ref{table_AG}.

{\sl Path 3}: The reentrant region and the approach to the attractive glass. This path is  studied with model A, at $\phi_c=0.58$, a value well within the
reentrant region\cite{Zac02a}. The corresponding  parameters for this path are provided  in Table
\ref{table_AG}. At large temperature, the glass transition is
approached but not reached because the studied packing fraction is
close, but smaller than $\phi_c^G$ for $V_{HS}$, i.e. the path is parallel
to the repulsive glass line in the limit $T\rightarrow \infty$.

Note that, as predicted from MCT, the attractive glass shows lower von
Schweidler exponents than the repulsive glass, for both paths and
models, while $\gamma_{\tau}$ is larger. This implies that the
divergence of the time scale for structural relaxation is more
abrupt. For the square well mixture, quantitative results from
simulations and MCT are available\cite{Sci03a}, predicting the
transition point at $\phi=0.58$ for $\Gamma^{G, MCT}\simeq 3.70$, in
quite good agreement with that estimated from the fits
$\Gamma^{G}\simeq 3.56$. For path 2 a quantitative comparison
with MCT has been also recently performed \cite{henrich07}, showing that the
driving mechanism for the slowing down observed in the simulation is
driven by the short-range attractions (large-$q$ modes of $S(q)$).

\begin{table}
\begin{tabular}{l|cccc}
 & $\Gamma^G$ & $b$ & $\gamma_{\tau}$ & $\gamma_D$ \\ \hline 
Path 2: $V_{tot}$ & $9.099$ & $0.37$ & $3.23$ & $1.23$  \\
Path 3: $V_{SW}$ & $3.56$ & $0.33$ & $3.75$ & $2.2$ \\
\end{tabular}
\caption{ Glass transition point $\Gamma^G$, von
Schweidler exponent $b$, and divergence exponents $\gamma_{\tau}$ and
$\gamma_D$ for models A and B in the presence of attraction, i.e. $V_{SW}$ and
$V_{tot}$, along respectively path 3 and 2.}
\label{table_AG}
\end{table}

\section{Viscosity results}

In this section we study the viscosity along the three paths described above. 

\subsection{Hard and soft spheres: Paths 1A and 1B}
\begin{figure}[h]    
\includegraphics[width=0.5\textwidth]{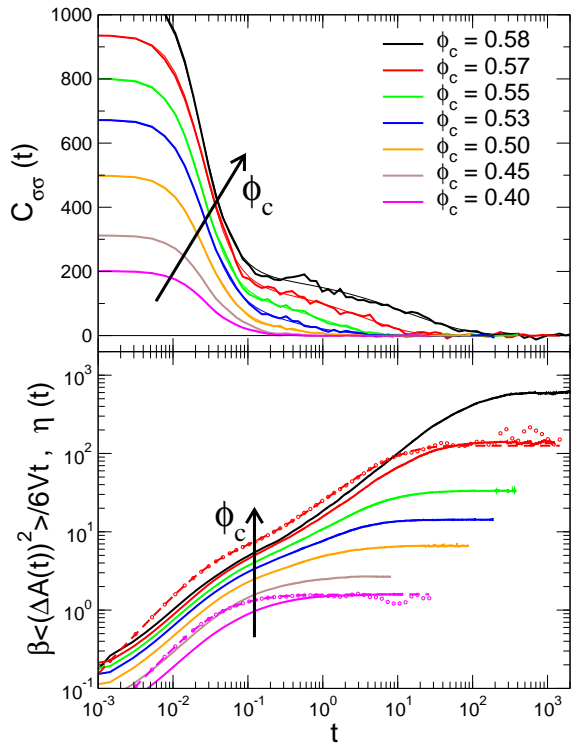}
\caption{\label{Css-HS} Upper panel: Stress correlation function
$C_{\sigma\sigma}(t)$ for $V_{sc}$. The thin lines  are empirical
fittings to describe the data (see section \ref{sec:MCT} for details).  
Lower panel:  Full lines are
 $\beta <(\Delta A(t))^2> /6Vt$  (from the Einstein relation  Eq.~\ref{eq:einstein}) 
for all studied $\phi_c$. For two
specific values of $\phi_c$ ($\phi_c=0.57$ and $\phi_c=0.40$) we also show 
$\eta(t)$ obtained using a direct integration of $C_{\sigma\sigma}(t)$ (symbols), and 
integration of the fitting curves (dashed thick). 
Note that while $\eta(t)$ and $\beta (\Delta A(t))^2/6Vt$ have the same 
long-time value, their time dependence is different.}
\end{figure}
In Figure \ref{Css-HS} we present, along path $1B$, the stress
correlation function for $V_{sc}$ at different concentrations (upper
panel), and the integral of the squared non-diagonal terms of the
stress tensor (lower panel). The correlation functions have been
averaged over $5000$ independent calculations. Note the progressive
development of a two-step decay in $C_{\sigma\sigma}(t)$ as the
concentration increases and the glass transition is approached, with
the second (structural) decay of $C_{\sigma\sigma}(t)$ moving to
longer and longer times. This implies that stress relaxes slower and
slower, or equivalently that the system increases its ability to store
the stress; i.e. the system becomes viscoelastic. Additionally, it can
be observed that $C_{\sigma\sigma}(0)$ grows close to the
transition. Both effects are responsible for the
increase of the viscosity upon increasing the packing fraction, but  the increase in the time scale is the one providing the leading contribution to the 
integral (see Eq.~\ref{eq:eta}). 

The integral of the stress correlation function is very noisy, and the
numerical evaluation of the viscosity is more accurate if calculated
using the Einstein relation (Eq.~\ref{eq:einstein}), as shown in the
lower panel of Fig.~\ref{Css-HS}.  For comparison, the integral of the
functional form used to describe $C_{\sigma\sigma}(t)$ (see below) is
also included for two state points. Note that all three quantities
show the same long-time limit, i.e. the viscosity does not depend on
the way it is calculated. At intermediate times, the integral of
$C_{\sigma\sigma}(t)$ and its fitting are in perfect agreement, but
the integral of the fitted function is less noisy. Thus, we will
calculate viscosities using the Einstein relation in
Eq.~\ref{eq:einstein}.

\begin{figure}[h]    
\includegraphics[width=0.5\textwidth]{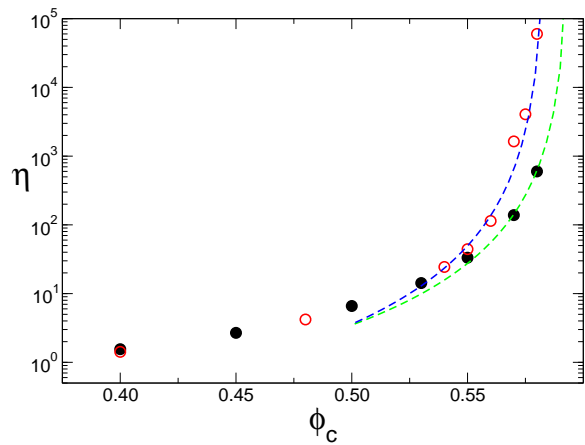}
\caption{Viscosity of soft (full black circles) and
hard (empty red circles) spheres as a function of particle packing fraction,
approaching the glass transition. Lines are power law
fits  to points with $\phi > 0.50$. The values of the critical packing fraction 
have been fixed to the previously determined values (see Table \ref{table_RG}), i.e.
 $\phi_c^G=0.594$ and $\phi_c^G=0.584$ for soft and hard sphere respectively.
 The corresponding fitting exponents $\gamma_{\eta}$ are $2.74$ and $2.9$.}
\label{viscosity-HS}
\end{figure}

The viscosity, as given by the long-time plateau, grows with
increasing particle density, as shown in Fig. \ref{viscosity-HS}. This
increase is consistent with a power-law, diverging at the transition point
estimated from the structural relaxation time and from the diffusion
coefficient, $\phi_c^G=0.594$ \cite{Pue05a}. The exponent for this
power-law $\gamma_{\eta}=2.74$ is similar to $\gamma_{\tau}$ but different from $\gamma_D$, as reported in Table \ref{table_RG}.

\begin{figure}[h]    
\includegraphics[width=0.5\textwidth]{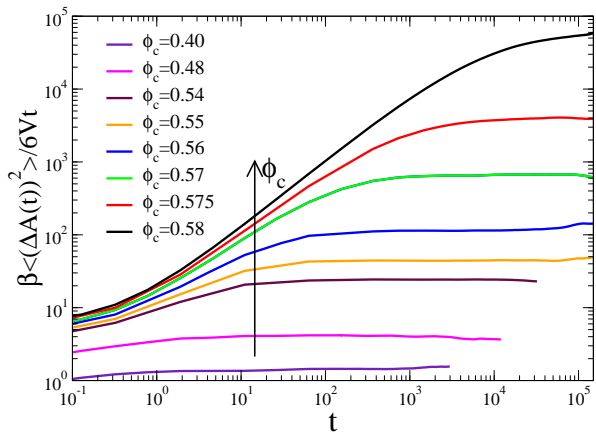}
\caption{\label{visc-HS} $\beta <(\Delta A(t))^2>/6Vt$ (with $\beta=1$) for hard
spheres, along path 1A.
}
\end{figure}

For hard spheres, path $1A$, we only show the integrated squared
non-diagonal terms --- obtained from Eq.\ref{eq:HSeta} --- in
Fig. \ref{visc-HS}. These results are obtained averaging over $20$
independent starting configurations and over time for a minimum of
$70\tau_{\alpha}$, where $\tau_{\alpha}$ is the density relaxation
time at the wavelength corresponding to the nearest-neighbour peak.
The behaviour of the curves is very similar to that shown above for
model B, and the viscosity, also shown in Fig. \ref{viscosity-HS},
increases as the glass transition is approached. A power-law
divergence with exponent $\gamma_{\eta}\simeq 2.9$ is observed for the
viscosity, with transition point at $\phi_c^G=0.584$, slightly lower
than for $V_{sc}$. The value of the exponent is, again, in good
agreement with $\gamma_{\tau}$ but quite different from $\gamma_D$.

\subsection{Attractive glass: Path $2$}

\begin{figure}[h]
\includegraphics[width=0.5\textwidth]{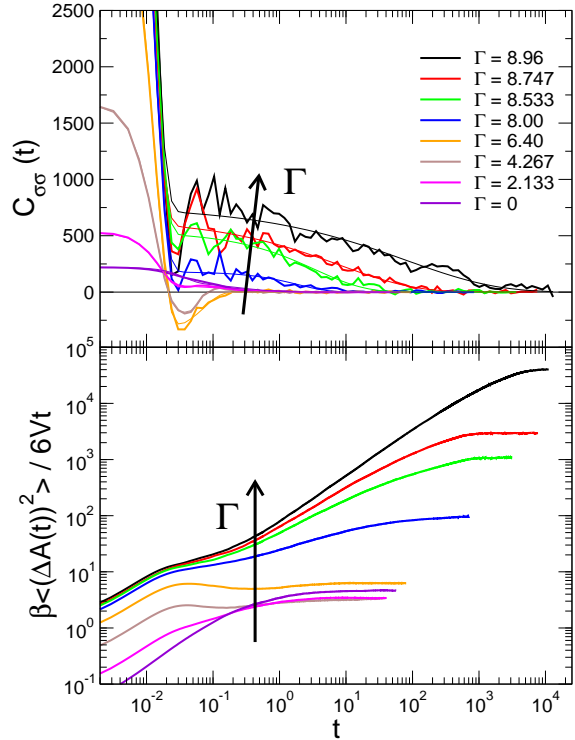}
\caption{ Stress correlation function
$C_{\sigma\sigma}(t)$ (upper panel) and $\beta<(\Delta A(t))^2>/6Vt$ (lower panel) 
for different state points along the isochore $\phi_c=0.40$. The thin
lines in the upper panel represent empirical fittings to $C_{\sigma\sigma}(t)$,
eq. \label{fit-AG} (see section \ref{sec:MCT} for details).}
\label{Css-AG}
\end{figure}

In this section, we analyse the viscoelastic behaviour close to the attractive
glass. As discussed above, for this purpose we use model $B$ for which 
the liquid-gas separation is suppressed by the presence of the added
repulsive barrier, allowing for the study of low
density ($\phi_c=0.40$) in a homogeneous system. In Fig. \ref{Css-AG},
 we present again
the stress correlation functions and the calculation of the viscosity
by integrating the squared stress tensor non-diagonal terms. The
attraction between particles induces a minimum after the short time
(microscopic) relaxation, which introduces a negative correlation at
intermediate attraction strengths. The origin of this minimum is
similar to that in the velocity auto-correlation function, although
here it is caused by stretching and rebound of the bonds. At high
attraction strength, the correlation is positive again at all times, 
and after the
minimum, $C_{\sigma\sigma}(t)$ shows the development of a two-step
decay and a large increase of the
value  at zero time $C_{\sigma\sigma}(0)$, similarly to the phenomenology observed
for the repulsive glass. This indicates that
the system is becoming solid-like.

\begin{figure}[h]    
\includegraphics[width=0.5\textwidth]{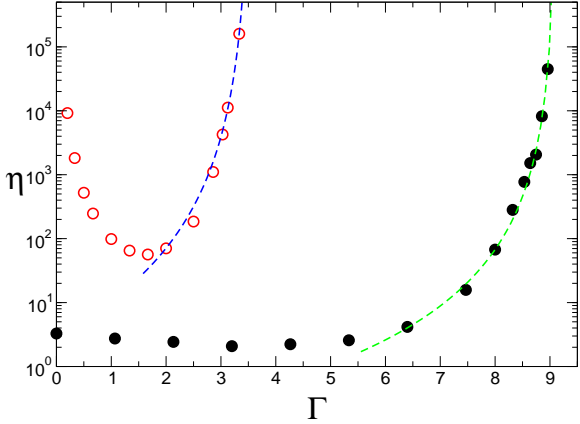}
\caption{ Viscosity 
approaching the attractive glass transition 
 along path $2$ (full black circles), and  in
the reentrant region along path $3$ (empty red circles), as a function of attraction strength. Lines
represent  power-law fittings (with values of the
critical attraction strength fixed to the previously determined values reported in Table~\ref{table_AG}), with exponents $\gamma_{\eta}$  equal to $3.16$ for path 2 and $3.75$ for the attractive side of the reentrant path 3.}
\label{viscosity-AG}
\end{figure}

$\langle (\Delta A(t))^2 \rangle$, shown in the lower panel of
Fig. \ref{Css-AG}, grows dramatically upon increasing the attraction
strength. The long time limit value, $\eta$, is shown in
Fig. \ref{viscosity-AG} as a function of attraction strength. The data
can be fitted using a power law divergence as a function of the
distance from the transition, $\Gamma -\Gamma^G$, where $\Gamma^G$ is
reported in Table~\ref{table_AG}. The exponent $\gamma_{\eta}=3.16$ is
again in good agreement with $\gamma_{\tau}$.

\subsection{Reentrance region: Path $3$}

As discussed above, path $3$ is a high density isochoric path, where
the attractive and repulsive glass lines are about to merge. Varying
the attraction strength, the system can be studied in states close to
the repulsive or to the attractive glass. This path is studied only
with system A, because the short interaction range of the studied SW
opens up a large fluid region between the two glasses.

\begin{figure}[h]
\includegraphics[width=0.5\textwidth]{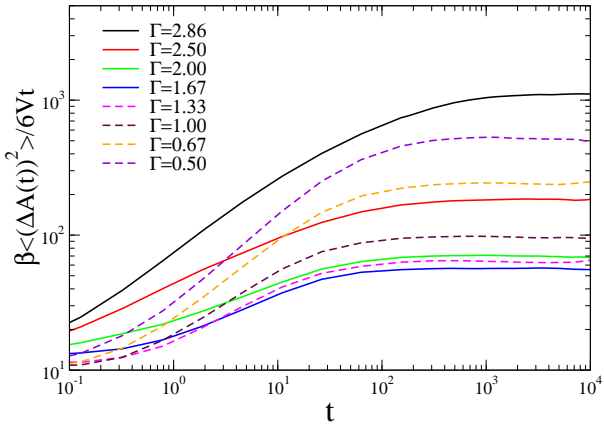}
\caption{$\beta <(\Delta A(t))^2>/6Vt$ for different attraction strength $\Gamma$
along the isochore $\phi_c=0.58$ for path 3A.  On decreasing $\Gamma$,
the long time limit first decreases (full lines) and then increases
again (dashed lines), resulting in a pronounced reentrant behaviour of
the viscosity.  
}
\label{visc-new}
\end{figure}

Fig.~\ref{visc-new} shows $\langle(\Delta A(t))^2\rangle/t$
calculated using Eq.\ref{eq:HSeta}. The corresponding viscosity is
reported in Fig. \ref{viscosity-AG} as a function of $\Gamma$. As
expected in this region, the viscosity increases both at low
temperature, due to the proximity of the attractive glass, and at high
temperature, because of the nearby repulsive glass. A power law
divergence describes the attractive glass increase of $\eta$ with
exponent $\gamma_{\eta}\simeq 3.75$, i.e. the same that is found also
for the density relaxation time $\gamma_{\tau}$.  Data refer to an
average over $20$ independent starting configurations and over time
for a minimum of $200\tau_{\alpha}$. A pronounced reentrant behaviour,
covering two full decades toward both limits, is observed in $\eta$,
similar to that reported previously for the diffusion coefficient $D$
in the same system\cite{Zac02a}. 

\section{Comparison of $C\sigma\sigma(t)$ with Mode Coupling Theory}
\label{sec:MCT}

MCT predicts\cite{Nag98b} that the stress correlation function is related to an
integral over all wavevectors of the density correlation functions:
\begin{equation}
C_{\sigma\sigma}(t)\:=\:\frac{k_BT}{60 \pi^2} \int_0^{\infty} dq\,q^4
\left[ \frac{d\,\ln S(q)}{dq}\,\Phi_q(t) \right]^2
\label{csigmasigmaMCT}
\end{equation}
We theoretically calculate $C_{\sigma\sigma}(t)$ along two paths analogous to paths
1B and 2 studied in simulations, to compare the full time-behaviour of
the stress correlation function. Hence, we study:

\noindent
(i) a one-component hard sphere system with increasing $\phi$,
using the Percus-Yevick (PY) structure factor as input;

\noindent
(ii) a one-component AO model with size ratio $q=0.1$ at fixed
packing fraction $\phi=0.40$. Here $S(q)$ is calculated using PY
closure for the two-component Asakura-Oosawa mixture. This model mixture
is composed of HS colloidal particles and ideal-gas 
polymers with HS interactions between polymers and
colloids\cite{Asa58a}. The obtained colloid-colloid structure factor 
is used as input to a one-component MCT, a treatment  based
on the validity of an effective one-component description for small polymer-colloid
size ratio\cite{Evans1,Evans}.  We did not use the fundamental measure
density functional theory \cite{Sch00bPRL,Sch02aJPCM} which yields analytical
expressions for $S_{ij}(k)$ as done previously\cite{Zac04a} because
within this closure the system shows spinodal instability before
MCT would actually give a glass. This is not the case with PY closure
for which only a very tiny increase in the structure factor at small $q$ is
found approaching the MCT transition.

We solved the full dynamical MCT equations, as well as their long
time limit, to calculate the viscoelastic properties close to the
glass transition. We used a grid a 1500 wave-vectors with mesh $\Delta q=0.314$.

\begin{figure}[tbh]    
\includegraphics[width=0.5\textwidth]{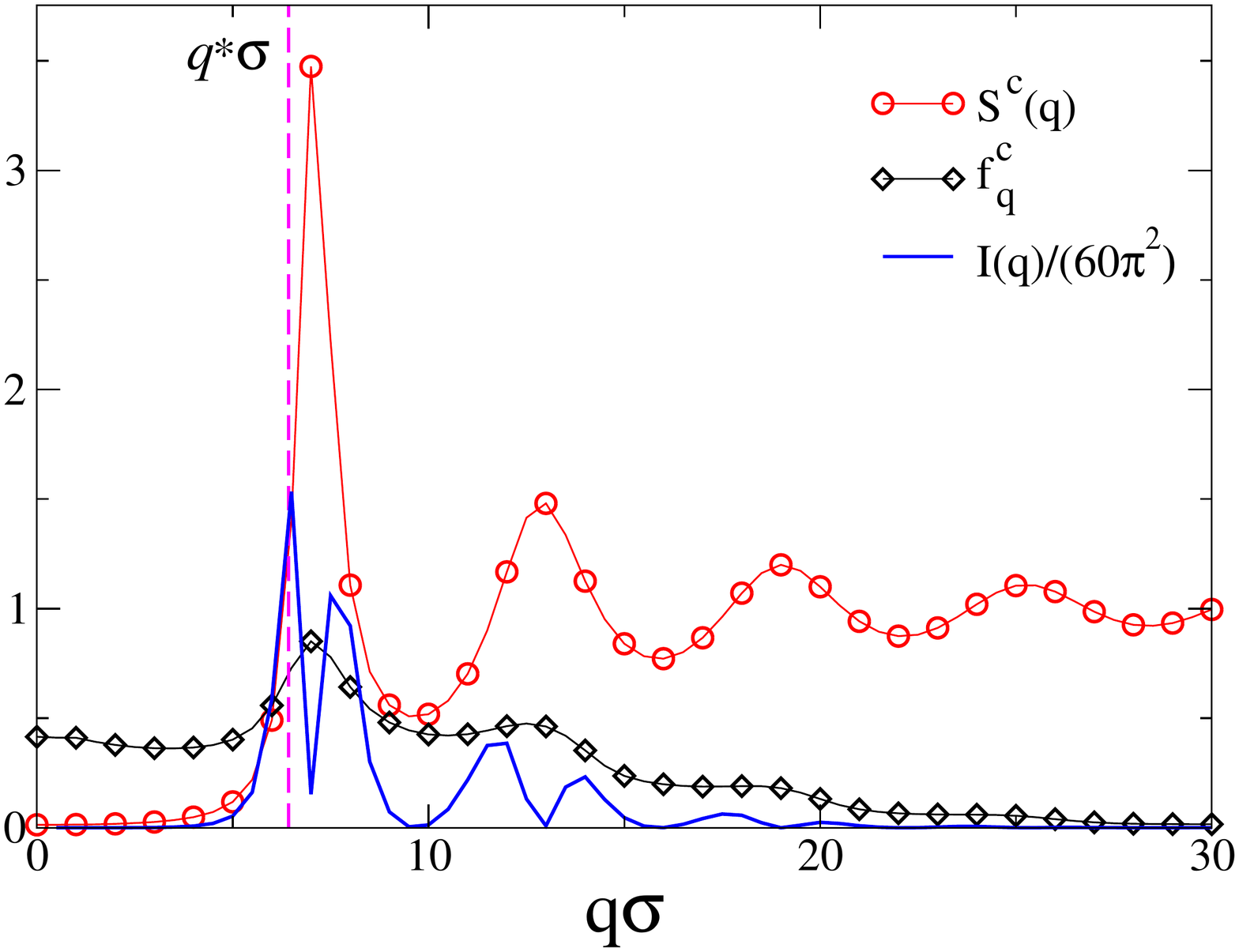}
\includegraphics[width=0.5\textwidth]{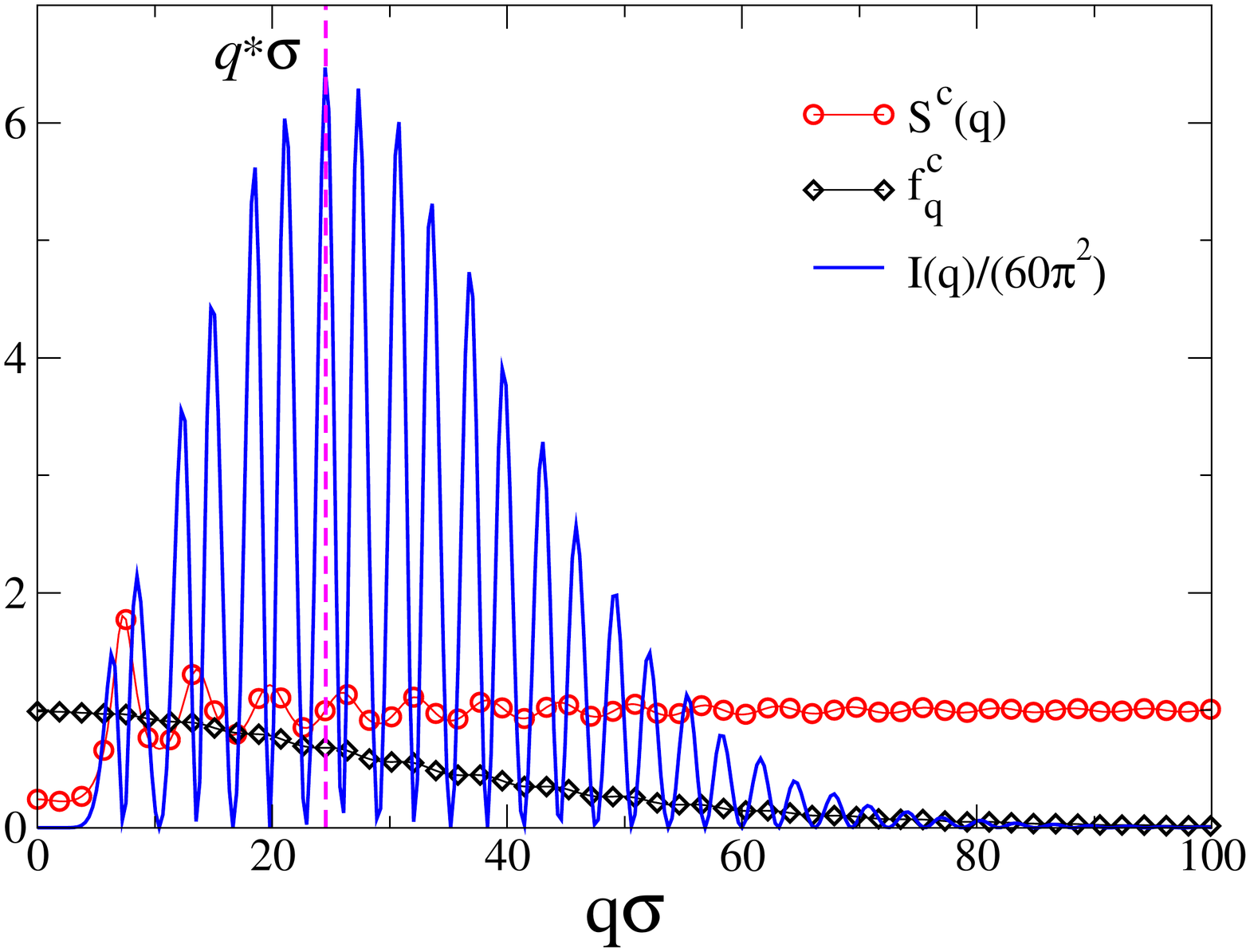}
\caption{Mode coupling contributions to the viscosity
$I(q)/(60\pi^2)$, with $I(q)$ defined in Eq.~\ref{eq:int}.  The wavevector at which 
$I(q)$ is maximum, $q^*\sigma$, is $\approx 6.5$ for the repulsive glass
and $\approx 24$ for the attractive glass. To compare, we report in the same figure also the $q-$dependence of the critical non-ergodicity parameter $f_q^c$ and of the static structure factor $S^c(q)$.}
\label{fig:MCT-q}
\end{figure}

The long-time limit of the integrand of Eq.~\ref{csigmasigmaMCT},
\begin{equation}
I(q)=\lim_{t\rightarrow\infty}q^4
\left[ \frac{d\,\ln S(q)}{dq}\,\Phi_q(t) \right]^2=
q^4 \left[ \frac{d\,\ln
S(q)}{dq}\,f^c_q\right]^2
\label{eq:int}
\end{equation}
is plotted as a function of $q\sigma$, 
in Figure \ref{fig:MCT-q} for both studied systems, $f^c_q$ being the
critical non-ergodicity parameter at the MCT transition. The same figure reports also 
 $f^c_q$ and the input static structure factor, also at the transition, $S^c(q)$.

For the repulsive glass we find that the dominant contribution to the
integral is provided by the wave-vector region around the nearest-neighbour peak, i.e. $q^*\sigma\approx 6.5$. 
For the attractive glass, on the other hand, the dominant contribution
is found at much larger $q$-values, i.e. $q^*\sigma\approx 24$ (in the
region of the fourth peak of $S(q)$) providing another confirmation of
the importance of small length-scales in the localization properties
of such a glass \cite{henrich07}. Moreover, in this case, the integrand is not just
peaked around a specific value, but it is rather spread within a very
large $q$-interval.  The amplitude of the integrand is also much
larger in the case of the attractive glass as compared to the
repulsive glass.

\begin{figure}[tbh]    
\includegraphics[width=0.5\textwidth]{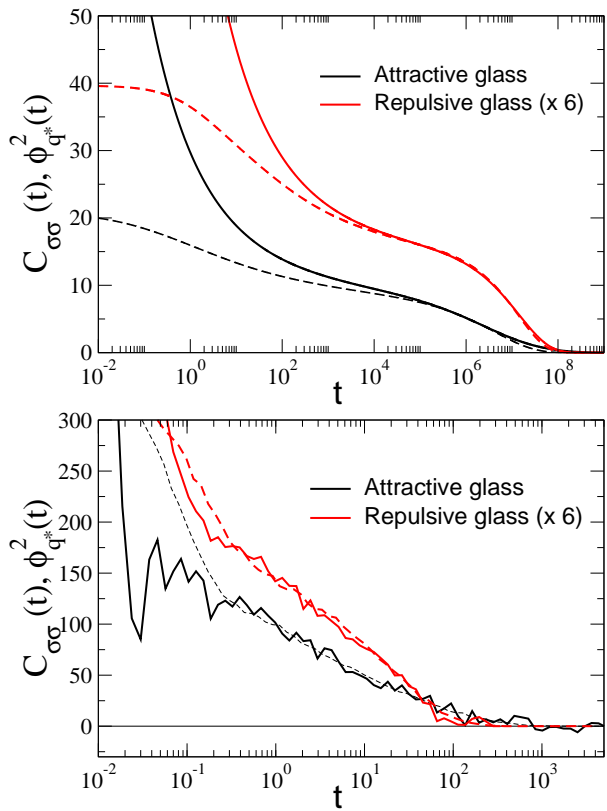}
\caption{Stress correlation function $C_{\sigma\sigma}(t)$ (full lines) for
repulsive and attractive glasses calculated within MCT (top) and from
simulations (bottom). Dashed lines are the squared density
correlation functions $\phi^2_{q^*}(t)$, arbitrarily scaled in amplitude to overlap the long time behavior.  For the MCT data, the wavevector $q^*$ is the one reported in Fig.~\ref{fig:MCT-q}, while in the simulation panel it is the one  which provides the best
long-time overlap between $\phi^2_{q^*}(t)$ and $C_{\sigma\sigma}(t)$. }
\label{figure20}
\end{figure}

We can then compare in the upper panel of Fig. \ref{figure20} the
theoretical stress correlation function with the squared theoretical 
density correlator $\phi^2_{q^*}(t)$ at the maximum of $I(q)$.
We show two state points, one close to the repulsive glass and the other
state close to the attractive one. Apart from an amplitude scaling factor, the dominant contribution is already sufficient to describe
the long-time behaviour of $C_{\sigma\sigma}(t)$ for both attractive and
repulsive glasses.  However, for the attractive glass  case, 
the decay of the squared density correlation shows a slightly smaller stretching as compared to $C_{\sigma\sigma}(t)$, which causes a small discrepancy at very long times. We attribute this difference to the fact that, in the case of attractive glasses, a large window of wavevectors contributes to the decay of the 
stress autocorrelation function (see Fig.~\ref{fig:MCT-q}). 

In the lower panel of Fig.~\ref{figure20}, the time dependence of both
$C_{\sigma\sigma}(t)$ and $\phi^2_{q^*}(t)$, as calculated from the
simulation data, are also plotted. Here $q^*$ is the wavevector at which the
agreement between the time dependence of $C_{\sigma\sigma}(t)$ and
$\phi^2_{q^*}(t)$ is optimal. The $q^*$ values found in this way,
respectively $q^*\sigma\approx 7.5$ and $q^*\sigma\approx 26$, agree
very well with those predicted by the theory\cite{Pue05a}. Moreover,
the behaviour of $C_{\sigma\sigma}(t)$ is well-described (within the
numerical error) by a single squared density correlator for both
glasses. The small discrepancy which was observed in the MCT data for the attractive glass is probably buried within the numerical noise.

Finally we want to compare the elastic moduli for both glasses in the
theoretical and numerical calculations.  In order to calculate elastic
and viscous moduli, the stress correlation functions calculated from
simulations have to be Fourier transformed: $G(\omega)=i\omega
\tilde{C}(\omega)$, where $\tilde{C}(\omega)$ is the Fourier transform
of $C_{\sigma\sigma}(t)$. However, due to the noise in the correlation
function, direct transformation produces very  low quality results. Thus, we
have fitted $C_{\sigma\sigma}(t)$ with empirical functional forms
close to both glasses before performing the Fourier transform. We have chosen 

\begin{eqnarray} \label{fit-Css}
C_{\sigma\sigma}(t)&=&C_{\sigma\sigma}(0) 
\left\{f(t/\tau_0)+\nonumber\right.\\
&&\left.A(1-f(t/\tau_0))\exp\{-(t/\tau_1)^{\beta}\}\right\}
\end{eqnarray}

\noindent where $f(x)$ is an even function that describes the short
time relaxation of $C_{\sigma\sigma}(t)$: $f(x)=1/(1+x^2)$ for the
repulsive glass (Fig. \ref{Css-HS}) and $f(x)=\exp\{-x^2\}$ for the
attractive glass (Fig. \ref{Css-AG}). $\tau_0$ represents a
microscopic time scale, which should be state-independent, whereas
$\tau_1$ gives the time scale for the stress final relaxation. The
parameter $A$ gives the amplitude of the stored stress (so that $AC_{\sigma\sigma}(0)$ is the height of
the plateau in $C_{\sigma\sigma}(t)$) and $\beta$ is the stretching
exponent, which according to the MCT prediction should be roughly
equal to the stretching exponent of the density-density correlation
function at $q^*$.

In Table \ref{table1} we present the parameters of the fittings for
$C_{\sigma\sigma}(t)$ for states along path $1B$, drawn in
Fig. \ref{Css-HS} as thin lines. As expected, $\tau_0$ is
state-independent and $\tau_1$ increases substantially when the glass
transition is approached. $A$ and $\beta$ are correctly estimated only
when the second relaxation is noticeable, i.e. above $\phi_c=0.55$; in
these cases the amplitude is almost constant and $\beta$ is compatible
with the value obtained from the density correlation function at $q^*$,
$\beta=0.52$ \cite{voightmann04}.

\begin{table}
\begin{tabular}{c|rrrrr}
$\phi_c$ & $C_{\sigma\sigma}(0)$ & $A$ & $\tau_0$ & $\tau_1$ & $\beta$ \\ \hline 
0.58 & 181 & 0.18 & 0.024 & 13.30 & 0.509 \\ 
0.57 & 156 & 0.16 & 0.026 & 3.56 & 0.665 \\ 
0.55 & 134 & 0.15 & 0.024 & 1.18 & 0.759 \\ 
0.53 & 83 & 0.23 & 0.025 & 0.20 & 0.421 \\ 
0.50 & 34 & 0.39 & 0.024 & 0.03 & 0.353
\end{tabular}
\caption{ Parameters of the fitting of $C_{\sigma\sigma}(t)$ for states close to glass transition for soft-spheres (path 1B).}
\label{table1}
\end{table}

The parameters of the fittings for the attractive glass (path 2), shown in
Fig. \ref{Css-AG}, are given in Table \ref{table2}. As
before, $\tau_0$ is almost constant, whereas $\tau_1$ increases
dramatically upon increasing the attraction strength.

\begin{table}[h]
\begin{tabular}{c|rrrrr}
$\phi_p$ & $C_{\sigma\sigma}(0)$ & $A$ & $\tau_0$ & $\tau_1$ & $\beta$ \\ \hline
0.42  &  1650  &  0.077  &  0.011  &  81.48  &  0.325  \\
0.41  &  1506  &  0.072  &  0.011  &  8.09  &  0.389  \\
0.40  &  1470  &  0.061  &  0.011  &  3.49  &  0.585  \\  
0.39  &  1404  &  0.071  &  0.012  &  1.90  &  0.949  \\
0.30  &  724 & -0.085 &  0.013  &  0.07  &  1.757  
\end{tabular} 
\caption{ Parameters of the fitting of
$C_{\sigma\sigma}(t)$ for states close to attractive glass
transition (Path 2).}
\label{table2}
\end{table}

\begin{table}[h]
\begin{tabular}{c|rrrr}
 & $C_{\sigma\sigma}(0)$ & $f_{\sigma}$ &  $C_{\sigma\sigma}^{MCT}(0)$ & $f_{\sigma}^{MCT}$\\ \hline
Path 1B  &   181  & 32  &  400  &  3  \\
Path 2   &  1650  &  127  &  6000  &  100  
\end{tabular} 
\caption{ Approximate values of initial value of the stress correlation
value $C_{\sigma\sigma}(0)$ and height of the plateau, $f_{\sigma}$
for paths 1B and 2. The first two columns refer to simulation data and the last two to 
theoretical   MCT predictions.}
\label{table_MCT}
\end{table}

From the values of the fits, we can directly compare other quantities
between theory and simulations: namely, the $t=0$ value of the stress
correlation function $C_{\sigma\sigma}(0)$ and the height of the
long-time plateau $f_{\sigma}$ for both glasses. The results from MCT
and simulations are reported in Table~\ref{table_MCT} for both studied
paths. For both glasses, the simulations provide a lower value of
$C_{\sigma\sigma}(0)$ and a larger value of $f_{\sigma}$ with respect
to MCT. Although numbers are not important {\it per se} when comparing
to MCT, the ratio $f_{\sigma}/C_{\sigma\sigma}(0)$ is wrong by one
order of magnitude for both attractive and repulsive glasses.  This
result seems to suggest that the factorization
approximation\cite{Nag98b} adopted to derive Eq.\ref{csigmasigmaMCT} may
be too severe, although the structural relaxation is apparently well described, as shown by the comparisons of Fig. \ref{figure20}.

\begin{figure}
\includegraphics[width=0.4\textwidth,clip]{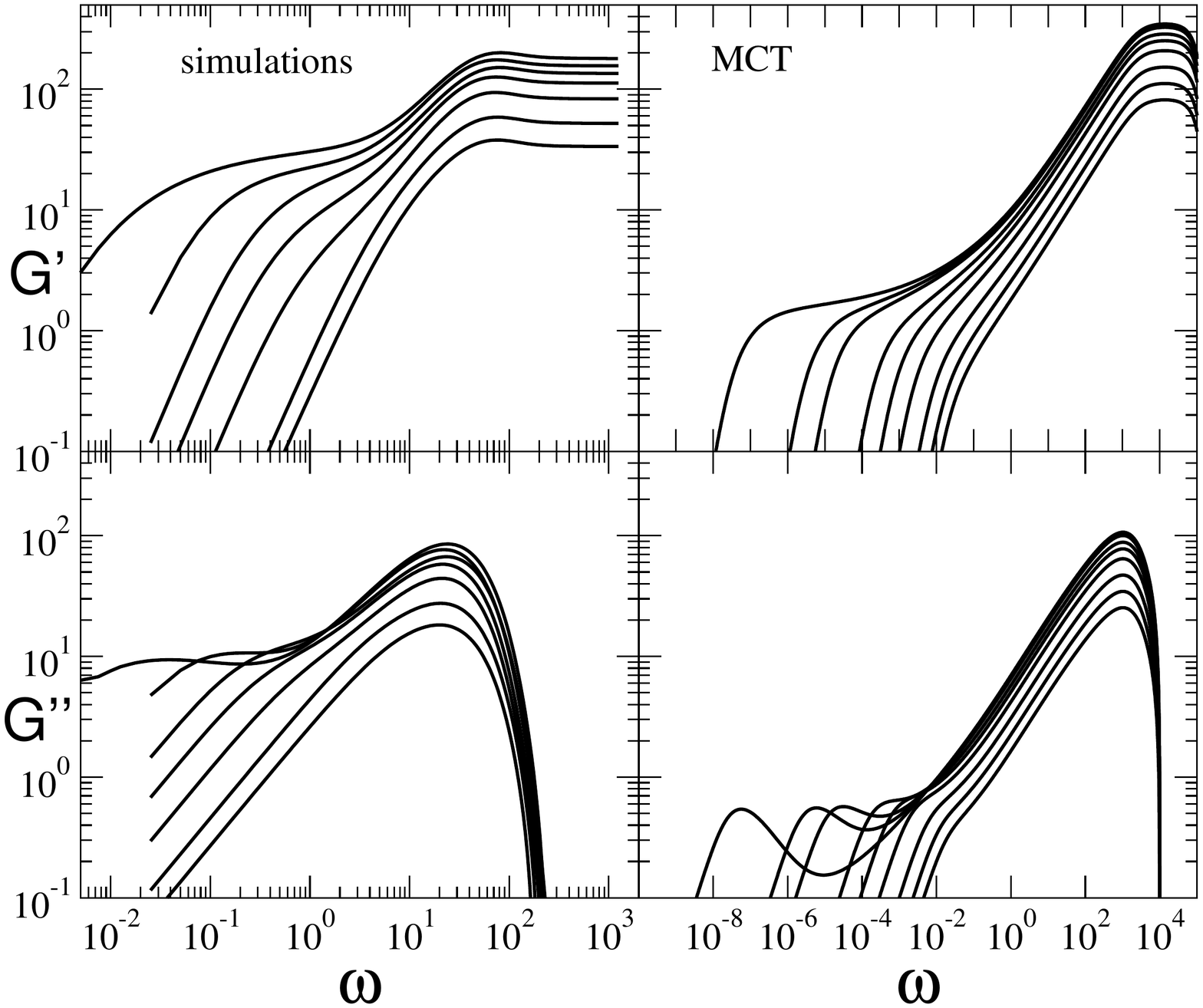}
\includegraphics[width=0.4\textwidth,clip]{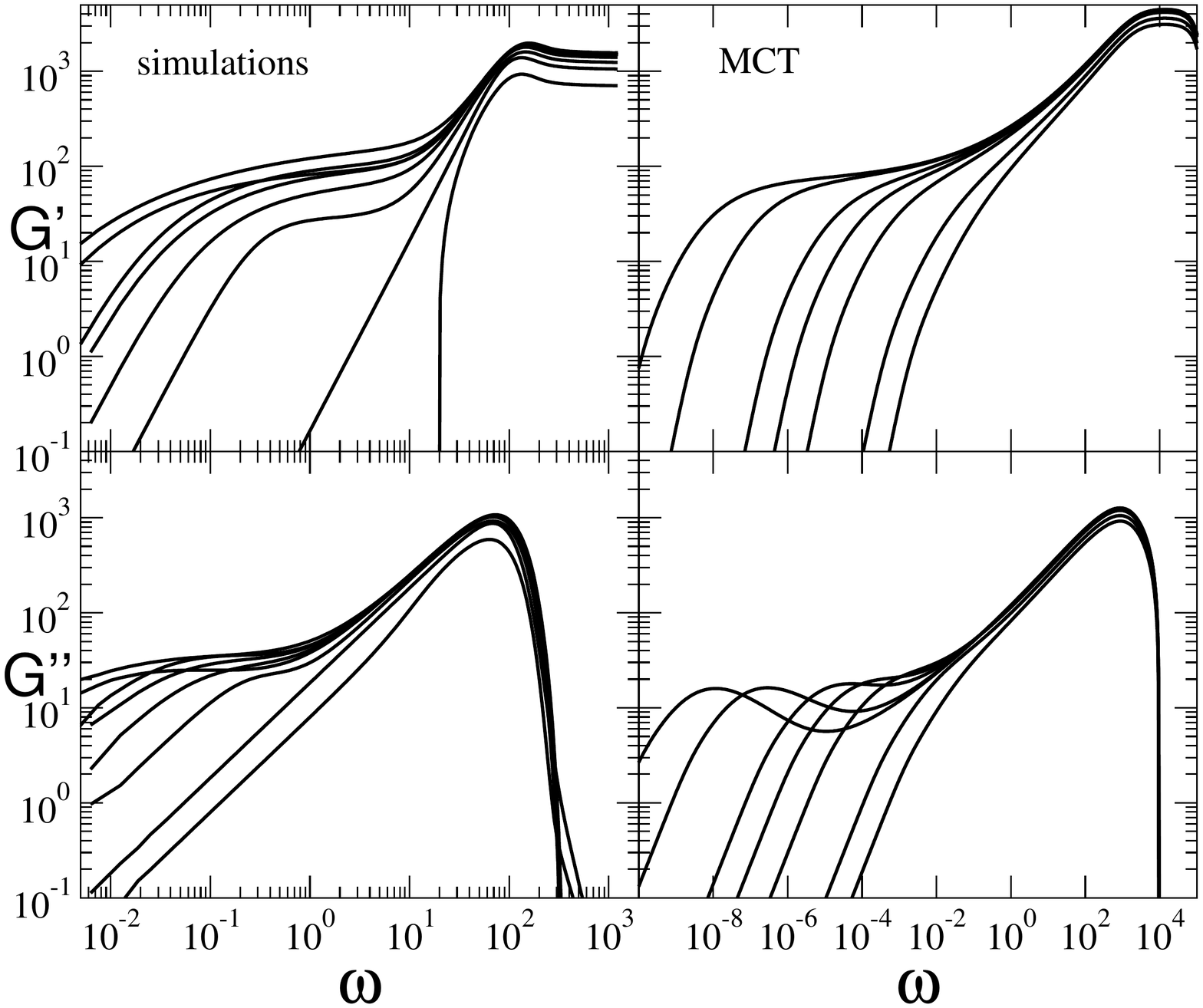}
\caption{Shear moduli $G'$ and $G''$ from simulations (left) and MCT
(right) for repulsive (top) and attractive glass (bottom).}
\label{moduli}
\end{figure}

We finally directly compare the elastic and viscous moduli
$G'(\omega)$ and $G''(\omega)$ in Fig.~\ref{moduli} for repulsive (top)
and attractive glass (bottom). We observe qualitatively the same
trends for both transitions in theory and simulations, despite a shift
in the absolute numbers:

(i) an increase of $G'(\omega)$ at  large-$\omega$ (but smaller than the microscopic frequency) with the approach to the glass transition;

(ii) the appearance of a minimum in $G''$ which moves to lower and
lower $\omega$ with decreasing distance from the transition, in agreement with previous experimental and theroetical studies on both repulsive  \cite{Mas95aPRL,Fuc99aPRE} and attractive glasses 
\cite{Daw01aJPCM,Mal04aJPCM}. The minimum 
appears when $\epsilon\lesssim0.01$ according to the theory 
($\epsilon=|X_g-X|/X_g$, with $X$ being either $\phi$ or $\Gamma$),
and at slightly larger values of $\epsilon$ according to the simulations;

(iii) much larger moduli (up to one order of magnitude) for the
attractive than for the repulsive glass. This observation holds both
for theory and simulations and agrees well with recent rheological
measurements for thermo-reversible sticky
spheres\cite{Nara2006PRL,Nara2006PRE}.

Overall, MCT correctly predicts the behavior of the viscoelastic
properties on approaching both glass transitions. However, the
results disagree again quantitatively, and more importantly in the
ratio of the height of the plateau in $G'$ (or minimum in $G''$) with
respect to $G'_{\infty}$ (or $G''_{max}$).

\section{Breakdown of Stokes-Einstein relation}

Finally, we discuss the breakdown of the Stokes-Einstein (SE)
relation\cite{Still94,ediger,kumar,Biroli07,garrahan,reichman,poole} close to the
glass transition for all different studied paths.

\begin{figure}    
\includegraphics[width=0.4\textwidth]{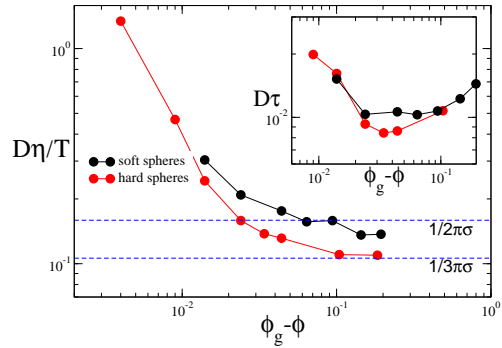}
\caption{Breakdown of the SE relation for $D\eta/T$ approaching the
repulsive glass transition for paths 1A (empty red circles) and 1B (full black circles). For the hard sphere case, $T=1$. Lines are guide to the eye. The two horizontal dashed
lines mark the slip and stick values of the SE relation. Inset: $D\tau$ for the same paths.
}
\label{D-eta-1}
\end{figure}

We start by examining path I.  Fig.~\ref{D-eta-1} shows the SE
relation for the hard sphere binary system and the soft sphere
polydisperse system.  To allow for a unifying picture, we plot the
results as a function of the relative distance to the estimated glass
transition $(\phi_g-\phi)$.  
At low and moderate density, far from the transition the data are consistent with SE, although different values limits are obtained for model A or B; whereas the former takes the stick value, $D\eta/T=(3\pi \sigma)^{-1}$, the latter goes to the slip limit: $D\eta/T=(2\pi \sigma)^{-1}$. The reason for this difference is not clear \cite{segre95,demichele01,moreno05}. In both cases, as the system approached the glass transition, the SE relation breaks down significantly, both in the form $D\eta$ and $D\tau$ (see inset).

\begin{figure}    
\includegraphics[width=0.4\textwidth]{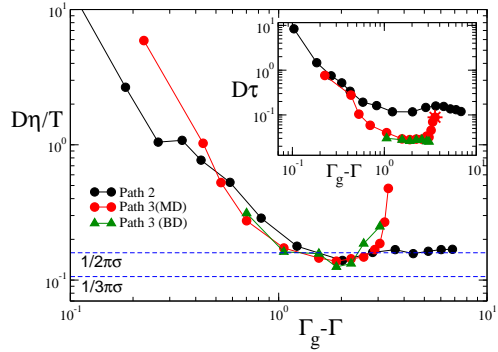}
\caption{Breakdown of the SE relation for $D\eta/T$ approaching the
attractive glass transition for paths 2 (circles) and 3 (squares-MD
and triangles-BD). Note the partial breakdown also at high $T$ for the
reentrant path due to the closeby repulsive glass. The two horizontal
lines mark the slip and stick SE values. Inset:  $D\tau$  for the same paths. The star indicates the HS value for path 3.
}
\label{D-eta-AG}
\end{figure}

Fig.~\ref{D-eta-AG} shows the SE relation for the attractive glass
case (path II) and along the reentrance (path III).  The former case
is rather clean, and allows us to access a breakdown by two orders of
magnitude with respect to the typical  SE value, both in $D\eta/T$ and $D\tau$
(inset).  For both paths, at large $\Gamma$ (low $T$) a clear
breakdown of both $D\tau$ and $D\eta/T$ is observed for the attractive
glass.

For path III (reentrance case), one has to bear in mind that the path
becomes parallel to the repulsive glass line at small $\Gamma$ (see
Fig.~1) and the increase is limited to the one observed in the HS case
at the same packing. For this path we have also performed BD simulations. The BD results, also shown in Fig.~\ref{D-eta-AG} coincide with the MD data at all state points investigated, confirming that the SE behavior close to
both repulsive and attractive glass transitions does not depend on the
microscopic dynamics.

Data in Fig.~\ref{D-eta-1} and Fig.~\ref{D-eta-AG} provide evidence
that the breakdown of the SE is a phenomenon which can be observed in
the vicinity of both the repulsive and the attractive glass
transitions.  Within the investigated state window, it appears
that the magnitude of the breakdown is enhanced in the attractive
glass case, speaking for the presence of more intense dynamical
heterogeneities \cite{Pue04b,Solomon06,Kilfoil} when confinement is
originated by short-range bonds rather than by the excluded volume
caging.
 
\section{Conclusions}
In this article we reported the behavior of the viscosity in two models
for short-range attractive colloids along three different paths in the
attraction-strength packing-fraction plane.  Along the first path, the
system approaches the repulsive hard-sphere glass transition. Along
the second path, it approaches the attractive glass. The third path is
chosen in such a way that the system moves continuously from the
repulsive to the attractive glass at constant packing fraction in the
so-called re-entrant region\cite{Sciortino2002b}.  In this case, we
have also compared brownian and newtonian simulation results,
confirming that the viscosity is independent on the microscopic
dynamics, in agreement with results based on the decay of density
fluctuations in atomic liquids\cite{Gle98}.

We find that the increase of the viscosity on approaching the glass
transition is consistent with a power-law divergence.  The divergence
of $\eta$ can be described with the same exponent and critical packing
fraction previously found for the collective relaxation time, but with
an exponent different from the one that characterizes the divergence of
the diffusion coefficient. This holds for both attractive and
repulsive glass.

As previously observed for diffusion and collective relaxation, the
viscosity shows a non monotonic behavior with the attraction strength
in the reentrant region (path III), confirming once more the validity
of the theoretical MCT predictions.

To provide a connection between density relaxation and visco-elastic
behavior we investigate the leading density fluctuation contributions
to the decay of the stress autocorrelation function within MCT. Interestingly,
for the case of the repulsive glass, it is possible to identify a
small range of wave-vectors (not far from the first peak of the
structure factor) which are responsible for the visco-elastic
behavior. In the case of the attractive glass, instead, the decay of
the stress is associated to a much larger window of wavevectors,
centered at much larger values. In this respect, the visco-elastic
analysis confirms that dynamic arrest is driven by the short-lengh
scale introduced by the bonding.  We also compare the simulation
results for the frequency dependence of the elastic moduli with
corresponding theoretical MCT predictions, finding a substantial
qualitative agreement.

Finally, we have evaluated the Stokes-Einstein relation. A clear
breakdown of the relation is observed on approaching both glass lines,
consistent with the different exponents characterizing the power-law
dependence of diffusion and viscosity.  The breakdown is particularly
striking on approaching the attractive glass (a variation of the
product $D\eta/T$ of up to two order of magnitude in the investigated
range). Recent theoretical work on MCT seems to provide insights that
could be useful to reconcile the decoupling of self-diffusion and
viscosity (or relaxation time) within MCT\cite{Biroli07}. It would be
interesting in the future to deepen our knowledge of the connection
between SE breakdown and the presence of dynamic heterogeneities,
which has been previously studied for the same model\cite{Pue04b}.

Note: While finalizing the manuscript, we become aware of a numerical
study by Krekelberg {\it et al.} (cond-mat/07050381) which also
reports the non-monotonic behavior of the viscosity along the
reentrant path and the breakdown of the SE relation. In that
work, Krekelberg {\it et al.} seek a connection between the structural
and dynamical properties of the system. We show here that MCT predicts
correctly the properties of the system upon approaching the glass
transitions, i.e. the connection between structure and dynamics is the
non-trivial one provided by MCT.

\section{Acknowledgments}
We thank M. Fuchs for stimulating discussions and S. Buldyrev for the
MD code.  We acknowledge support from MIUR-Prin and
MRTN-CT-2003-504712. A.M.P. was financially supported by the Spanish
Ministerio de Educaci\'on y Ciencia (under Project
No. MAT2006-13646-CO3-02).

\bibliographystyle{./apsrev}
\bibliography{./visco-final}

\end{document}